\documentstyle[12pt]{article}
\newfont{\ffont}{msym10}                        
\newcommand{\beq}{\begin{equation}}             
\newcommand{\eeq}{\end{equation}}               
\newcommand{\bqry}{\begin{eqnarray}}            
\newcommand{\eqry}{\end{eqnarray}}              
\newcommand{\bqryn}{\begin{eqnarray*}}          
\newcommand{\eqryn}{\end{eqnarray*}}            
\newcommand{\NL}{\nonumber \\}                  
\newcommand{\preprint}[1]{\begin{table}[t]      
            \begin{flushright}                  
            \begin{large}{#1}\end{large}        
            \end{flushright}                    
            \end{table}}                        
\newcommand{\PD}[2]                             
    {\frac{\partial^{#2}}{\partial #1^{#2}}}    
\catcode`\@=11 \@addtoreset{equation}{section}  
\renewcommand{\theequation}                     
         {\arabic{section}.\arabic{equation}}   
\begin{document}
\preprint{TAUP-2149-94 \\ LA-UR-96-XX \\ IAS-SNS-96/32 }
\title{A New Relativistic High Temperature \\ Bose-Einstein Condensation}
\author{\\ L. Burakovsky\thanks{Bitnet: BURAKOV@QCD.LANL.GOV} \
\\  \\  Theoretical Division, T-8 \\  Los Alamos National Laboratory \\ Los
Alamos NM 87545, USA
 \\  \\
L.P. Horwitz\thanks{Bitnet: HORWITZ@SNS.IAS.EDU. On sabbatical leave from
School of Physics and Astronomy, Tel Aviv University, Ramat Aviv, Israel.
Also at Department of Physics, Bar-Ilan University, Ramat-Gan, Israel  } \
\\  \\
School of Natural Sciences \\ Institute for Advanced Study \\ Princeton NJ
 08540, USA \\  \\ and \\  \\ W.C. Schieve\thanks{Bitnet: WCS@MAIL.UTEXAS.EDU}\
\\  \\ Ilya Prigogine Center \\ for Studies in Statistical Mechanics \\
University of Texas at Austin \\ Austin TX 78712, USA \\}
\date{ }
\maketitle
\begin{abstract}
We discuss the properties of an ideal relativistic gas of events
possessing Bose-Einstein statistics. We find that the mass spectrum of
such a system is bounded by $\mu \leq m\leq 2M/\mu _K,$ where $\mu $ is
the usual chemical potential, $M$ is an intrinsic dimensional scale
parameter for the motion of an event in space-time, and $\mu _K$ is an
additional mass potential of the ensemble. For the system including both
particles and antiparticles, with nonzero chemical potential $\mu ,$ the
mass spectrum is shown to be bounded by $|\mu |\leq m\leq 2M/\mu _K,$ and
a special type of high-temperature Bose-Einstein condensation can occur.
We study this Bose-Einstein condensation, and show that it corresponds to
a phase transition from the sector of continuous relativistic mass
distributions to a sector in which the boson mass distribution becomes
sharp at a definite mass $M/\mu _K.$ This phenomenon provides a mechanism
for the mass distribution of the particles to be sharp at some definite
value.
\end{abstract}
\bigskip
{\it Key words:} special relativity, relativistic Bose-Einstein
condensation, mass distribution, mass shell

PACS: 03.30.+p, 05.20.Gg, 05.30.Ch, 98.20.--d
\bigskip
\section{Introduction}
There have been a number of papers in the past \cite{Al,BKM1,CR,BKM2},
which discuss the properties of an ideal relativistic Bose gas with
nonzero chemical potential $\mu .$ Particular attention has been given to
the behavior of the Bose-Einstein condensation and the nature of the
phase transition in $d$ space dimensions \cite{BKM2,Lan}. The basic work
was done many years ago by J\"{u}ttner \cite{Jut},
Glaser \cite{Gla}, and more recently by Landsberg and Dunning-Davies
\cite{LDD} and Nieto \cite{Nie}. These works were all done in the
framework of the usual on-shell relativistic statistical mechanics.

To describe an ideal Bose gas in the grand canonical ensemble, the
usual expression for the number of bosons $N$ in relativistic
statistical mechanics is
\beq
N=V\sum _{{\bf k}}n_k=V\sum _{{\bf k}}\frac{1}{e^{(E_k-\mu )/T}-1},
\eeq
where $V$ is the system's three-volume, $E_k=\sqrt{{\bf k}^2+m^2}$ and
$T$ is the absolute temperature (we use the system of units in which
$\hbar =c=k_B=1;$ we also use the metric $g^{\mu \nu }=(-,+,+,+)),$ and
one must require that $\mu \leq m$ in order to ensure a
positive-definite value for $n_k,$ the number of bosons with momentum
${\bf k}.$ Here $N$ is assumed to be a conserved quantity, so that it
makes sense to talk of a box of $N$ bosons. This can no longer be true
once $T\stackrel{>}{\sim }m$ [10]; at such temperatures quantum field
theory requires consideration of particle-antiparticle pair production.
If $\bar{N}$ is the number of antiparticles, then $N$ and $\bar{N}$ by
themselves are not conserved but $N-\bar{N}$ is. Therefore, the
high-temperature limit of (1.1) is not relevant in realistic physical
systems.

The introduction of antiparticles into the theory in a systematic way was
made by Haber and Weldon \cite {HW1,HW2}. They considered an ideal Bose
gas with a conserved quantum number (referred to as ``charge'') $Q,$
which corresponds to a quantum mechanical particle number operator
commuting with the Hamiltonian $H.$\footnote{In the manifestly covariant
theory which we shall use in our study, this charge is naturally associated
with particles and antiparticles which are distinguished by the off-shell
structure, as in quantum field theory \cite{HW1}.} All thermodynamic quantities
may be then obtained from the grand partition function $Tr\;\{\exp \;[-(
H-\mu Q)/T]\}$ considered as a function of $T,V,$ and $\mu $
\cite{Hua}. The formula for the conserved net charge, which replaces
(1.1), reads\footnote{ The standard recipe according to which all
additive thermodynamic quantities are reversed for antiparticles is used.} [10]
\beq
Q=V\sum _{{\bf k}}\left[\frac{1}{e^{(E_k-\mu )/T}-1}-
\frac{1}{e^{(E_k+\mu )/T}-1}\right].
\eeq
In such a formulation a boson-antiboson system is described by only one
chemical potential $\mu ;$ the sign of $\mu $ indicates whether particles
outnumber antiparticles or vice versa. The requirement that both $n_k$
and $\bar{n}_k$ be positive definite leads to the important relation
\beq
|\mu |\leq m.
\eeq
The sum over ${\bf k}$ in (1.2) can be replaced by an
integral, so that the charge density $\rho \equiv Q/V$ becomes
\beq
\rho =\frac{1}{2\pi ^2}\int _0^\infty k^2\;dk\left[
\frac{1}{e^{(E_k-\mu )/T}-1}-(\mu \rightarrow -\mu )\right],
\eeq
which is an implicit formula for $\mu $ as a function of $\rho $ and $T,$
and in the region $T>>m$ reduces to
\beq
\rho \cong \frac{\mu T^2}{3}.
\eeq
For $T$ above some critical temperature $T_c,$ one can always find a $\mu
$ ($|\mu |\leq m)$ such that (1.4) holds. Below $T_c$ no such $\mu $ can
be found, and (1.4) should be interpreted as the charge density of the
excited states: $\rho -\rho _0,$ where $\rho _0$ is the charge density of
the ground state [10] (with ${\bf k}=0;$ clearly, this state is given
with zero weight in the integral (1.4)). The critical temperature $T_c$
at which Bose-Einstein condensation occurs corresponds to $\mu =\pm m$
(depending on the sign of $\rho .)$ Thus, one sets $|\mu |=m$ in (1.4)
and obtains, via (1.5) (provided that $|\rho |>>m^3),$
\beq
T_c=\sqrt{\frac{3|\rho |}{m}}.
\eeq
Below $T_c,$ (1.4) is an equation for $\rho -\rho _0,$ so that the charge
density in the ground state is
\beq
\rho _0=\rho [1-(T/T_c)^2].
\eeq
It follows from Eq. (1.6) that any ideal Bose gas will condense at a
relativistic temperature ($T_c>>m),$ provided that $|\rho |>>m^3.$

Recently an analogous phenomenon has been studied in relativistic
quantum field theory \cite{HW2,Kap,BD,BBD}. For relativistic fields
Bose-Einstein condensation occurs at high temperatures and can be
interpreted in terms of a spontaneous symmetry breaking \cite{HW2}.

In this paper we shall use a manifestly covariant form of statistical
mechanics which has  more general structure than the standard forms
of  relativistic statistical mechanics, but which reduces to those theories
in a certain limit, to be described precisely below. In fact, it is one of the
principle aims of this work to  provide a mechanism for which  this limit can
be realized on a statistical level. The results that we obtain are
 different from those of the standard theories at high  temperatures.
These theories, which are characterized classically by mass-shell
 constraints, and the use, in quantum field theory, of fields which are
 constructed on the basis of on-mass-shell free fields, are associated
with the statistical treatment of {\it world lines} and hence,
 considerable coherence (in terms of the macroscopic structure of
 whole world lines as the elementary objects of the theory) is implied.
 In nonrelativistic statistical mechanics, the elementary objects of the
theory are points. The relativistic  analog of this essentially structureless
 foundation for a statistical theory is the set of points in spacetime, i.e.,
 the so-called {\it events}, not the world lines (Currie,
 Jordan and Sudarshan \cite {Sudarshan} have discussed the
 difficulty of constructing a relativistic mechanics on the
 basis of world lines).

The mass of particles in a mechanical theory of events is necessarily
 a dynamical variable, since the classical phase space of
the relativistic set of events consists of the spacetime
 and energy-momentum coordinates $\{{\bf q}_i, t_i;\  {\bf p}_i,E_i\}$,
 with no {\it a priori} constraint on the relation between
the ${\bf p}_i$ and the $E_i$, and hence  such theories are
``off-shell''. It is well known from the work of Newton and Wigner
 \cite{NW} that on-shell relativistic quantum theories such as those
 governed by Klein-Gordon or Dirac type equations do not provide local
 descriptions (the wave functions corresponding to localized particles
 are  spread out); for such theories the notion of ensembles over local
 initial conditions is difficult to formulate. The off-shell theory that
 we shall use here is,  however, precisely local in both its first and
 second quantized forms \cite{AH,Shnerb}.

The phenomenologial predictions of on-shell theories, furthermore, provide
 equations of state which appear to be too rigid. Shuryak \cite{Shuryak} has
 obtained equations of state which are more realistic by taking into account
 the spectrum of mass as seen in the resonance spectrum of strongly
 interacting matter. We have shown \cite{Realistic} that Shuryak's
 ``realistic'' equation of state follows in a natural way from the
 mass distribution functions of the off-shell theory.

We finally remark that the standard formulations of  quantum relativistic
 statistical mechanics, and quantum field theory at finite temperature, lack
manifest covariance on a fundamental level. As for nonrelativistic statistical
mechanics, the partition function is described by the Hamiltonian, which is not
an invariant object, and hence thermodynamic mean values do not have tensor
properties. [One could consider the invariant $p_\mu n^\mu$ in place of the
Hamiltonian \cite{Synge}, where $n^\mu$ is a unit four-vector; this construction
(supplemented by a spacelike vector othogonal to $n^\mu$) implies an induced
representation for spacetime. The quantity that takes the place of the parameter
$t$ is then $x_\mu n^\mu$. This construction is closely related to the problem
pointed out by Currie, Jordan and Sudarshan \cite{Sudarshan}, for which
different world lines are predicted dynamically by the change in the form of the
effective Hamiltonian in different frames.] Since the form of such a theory is
not constrained by covariance requirments, its dynamical structure
 and
predictions may be different than for a theory which satisifies these
requirements. For example,  the canonical distribution of Pauli \cite{Pauli}
for the free Boltzmann gas has a high temperature limit in which the energy is
given by $3k_BT$, which does not correspond to  any known equipartition
rule, but
for the corresponding distribution for the manifestly covariant theory, the
limit is $2k_BT$, corresponding to ${1\over 2}k_BT$ for each of the four
relativistic degrees of freedom.  For the quantum field theories at finite
temperature, the path integral formulation \cite {Kap1} replaces the Hamiltonian
in the canonical exponent by the Lagrangian due to the infinite product of
factors $\langle\phi \vert \pi\rangle$ (transition matrix element of the
canonical field and its conjugate required to give a Weyl ordered Hamiltonian
its numerical value). However, it is the $t$ variable which is analytically
continued to construct the finite temperature canonical ensemble, completely
removing the covariance of the theoretical framework.  One may argue that some
frame has to be chosen for the statistical theory to be developed, and perhaps
even for temperature to have a meaning, but as we have remarked above, the
requirement of relativistic covariance has dynamical consequences (note that the
model Lagrangians used in the non-covariant formulations are established with
the criterion of relativistic covariance in mind), and we argue that the choice
of a frame, if necessary for some physical reason, such as the definition  and
measurement of temperature, should be made in the framework of a manifestly
covariant  structure.

We consider, in this paper, a relativistic Bose gas within the framework
 of a manifestly covariant relativistic statistical mechanics  \cite{HSP,HSS,BH1}.
We obtain the expressions for characteristic thermodynamic quantities and show
that they coincide quantitatively, in the narrow mass-width approximation, with
those of the relativistic on-shell theory, except for the value of the average
energy (which differs by a factor $2/3$, as remarked above). We introduce
antiparticles and discuss the high temperature Bose-Einstein condensation in
such a particle-antiparticle system. We show that it corresponds to a phase
transition to a high-temperature form of the usual on-shell relativistic kinetic
theory. In the following, we briefly review the manifestly covariant mechanics
and quantum mechanics which forms the basis of our study of relativistic
statistical mechanics.

In the framework of a manifestly covariant relativistic statistical
mechanics, the dynamical evolution of a system of $N$ particles, for the
classical case, is governed by equations of motion that are of the form
of Hamilton equations for the motion of $N$ $events$ which generate the
space-time trajectories (particle world lines) as functions of a
continuous Poincar\'{e}-invariant parameter $\tau ,$ called the
``historical time''\cite{Stu,HP}. These events are characterized by
their positions $q^\mu =(t,{\bf q})$ and energy-momenta $p^\mu =(
E,{\bf p})$ in an $8N$-dimensional phase-space. For the quantum case, the
system is characterized by the wave function $\psi _\tau (q_1,q_2,\ldots
,q_N)\in L ^2(R^{4N}),$ with the measure $d^4q_1d^4q_2\cdots d^4q_N\equiv
d^{4N}q,$ $(q_i\equiv q_i^\mu ;\;\;\mu =0,1,2,3;\;\;i=1,2,\ldots ,N),$
describing the distribution of events, which evolves with a generalized
Schr\"{o}dinger equation \cite{HP}. The collection of events (called
``concatenation'' \cite{AHL}) along each world line corresponds to a
{\it particle,} and hence, the evolution of the state of the $N$-event
system describes, {\it a posteriori,} the history in space and time of
an $N$-particle system.

For a system of $N$ interacting events (and hence, particles) one takes
\cite{HP}
\beq
K=\sum _i\frac{p_i^\mu p_{i\mu }}{2M}+V(q_1,q_2,\ldots ,q_N),
\eeq
where $M$ is a given fixed parameter (an intrinsic property of the
particles), with the dimension of mass, taken to be the same for all the
particles of the system. The Hamilton equations are
$$\frac{dq_i^\mu }{d\tau }=\frac{\partial K}{\partial p_{i\mu }}=\frac{p_
i^\mu }{M},$$
\beq
\frac{dp_i^\mu }{d\tau }=-\frac{\partial K}{\partial q_{i\mu }}=-\frac{
\partial V}{\partial q_{i\mu }}.
\eeq
In the quantum theory, the generalized Schr\"{o}dinger equation
\beq
i\frac{\partial }{\partial \tau }\psi _\tau (q_1,q_2,\ldots ,q_N)=K
\psi _\tau (q_1,q_2,\ldots ,q_N)
\eeq
describes the evolution of the $N$-body wave function
$\psi _\tau (q_1,q_2,\ldots ,q_N).$ To illustrate the meaning of this
wave function, consider the case of a single free event. In this case
(1.10) has the formal solution
\beq
\psi _\tau (q)=(e^{-iK_0\tau }\psi _0)(q)
\eeq
for the evolution of the free wave packet. Let us represent $\psi _\tau
(q)$ by its Fourier transform, in the energy-momentum space:
\beq
\psi _\tau (q)=\frac{1}{(2\pi )^2}\int d^4pe^{-i\frac{p^2}{2M}\tau }
e^{ip\cdot q}\psi _0(p),
\eeq
where $p^2\equiv p^\mu p_\mu ,$ $p\cdot q\equiv p^\mu q_\mu ,$ and $\psi
_0(p)$ corresponds to the initial state. Applying the Ehrenfest arguments
of stationary phase to obtain the principal contribution to $\psi _\tau
(q)$ for a wave packet at $p_c^\mu ,$ one finds ($p_c^\mu $ is the peak
value in the distribution $\psi _0(p))$
\beq
q_c^\mu \simeq \frac{p_c^\mu }{M}\tau ,
\eeq
consistent with the classical equations (1.9). Therefore,
the central peak of the wave packet moves along the classical
trajectory of an event, i.e., the classical world line.

In the case that $p_c^0=E_c<0,$ we see, as in Stueckelberg's classical
example \cite{Stu}, that $$\frac{dt_c}{d\tau }\simeq \frac{E_c}{M}<0.$$
It has been shown \cite{AHL} in the analysis of an evolution operator
with minimal electromagnetic interaction, of the form $$K=\frac{(p-eA(q)
)^2}{2M},$$ that the $CPT$-conjugate wave function is given by
\beq
\psi _\tau ^{CPT}(t,{\bf q})=\psi _\tau (-t,-{\bf q}),
\eeq
with $e\rightarrow -e.$ For the free wave packet, one has
\beq
\psi _\tau ^{CPT}(q)=\frac{1}{(2\pi )^2}\int d^4p
e^{-i\frac{p^2}{2M}\tau }e^{-ip\cdot q}\psi _0(p).
\eeq
The Ehrenfest motion in this case is $$q_c^\mu \simeq -\frac{p_c^\mu }{M}
\tau ;$$ if $E_c<0,$ we see that the motion of the event in the
$CPT$-conjugate state is in the positive direction of time, i.e.,
\beq
\frac{dt_c}{d\tau }\simeq -\frac{E_c}{M}=\frac{|E_c|}{M},
\eeq
and one obtains the representation of a positive energy generic event
with the opposite sign of charge, i.e., the antiparticle.

It is clear from the form of  (1.10) that one can construct  relativistic
 transport theory in a form analogous to that of the nonrelativistic theory; a
relativistic Boltzmann equation and its consequences, for example, was studied
in ref. \cite{HSS}.

As a simple example of the implications of the classical dynamical equations
 (1.9), consider the problem of a relativistic particle in a uniform  external
``gravitational'' field, with evolution  function
\beq
K=  {p_\mu p^\mu \over 2M }+ Mgz
\eeq
(the external potential breaks the invariance of the  evolution function,
 but that will not affect  the illustrative value of the example) with initial
conditions  $t(0) = 0, \dot{t}(0) = \alpha, z(0)= h, \dot{z}(0) = 0$, resulting
in the solution
\bqry
z & = & - {1 \over 2} g \tau^2 + h, \;\;\; t \;\;=\;\; \alpha \tau + t_0, \NL
E & = & Mc^2 \alpha, \;\;\;\;\;\;\;\;\;\; p_z \; = \; - Mg\tau .
\eqry
The invariant variable $\tau $ replaces $t$ in describing the dynamical
 evolution of the system. The generator of the  motion
\beq
K = {p_z^2 - E^2/c^2 \over 2M} +mgz = {1 \over 2} Mc^2 \alpha^2 = {\rm const,}
\eeq
as required. The total energy of the particle in this case, including
 {\it both} increase of momentum and decrease of dynamical mass, is constant
also. The effective particle mass $\tilde m$ is given by
\beq
\tilde m = {1 \over c} \sqrt{(E/c)^2 -p_z^2 } = M\alpha
 \sqrt{1- {g^2\tau ^2 \over c^2\alpha^2}}.
\eeq
Expanding this out in the nonrelativistic limit $c \rightarrow
 \infty $, one obtains (with $\tau^2 = 2(h-z)/g$)
\beq
\tilde m \cong M\alpha - {Mg \over \alpha c^2}(h-z),
\eeq
and we recognize $Mg(h-z)/c^2$ as the mass shift induced by the potential
term. The factor $\alpha$ arises due to the choice of initial conditions, i.e.,
for $\tau = 0, \tilde m = M\alpha$, and not $M$ (for $\tau$ sufficiently large,
under this unbounded potential, the quantity in the square root could become
negative, and the particle could become tachyonic). Note that it is the {\it
mass} of the particle which carries dynamical information (the total energy is
constant, but the mass is ``redshifted'' by the potential) and that has the
correspondence with nonrelativistic energy, through the mass-energy equivalence,
that we observe in the laboratory. This point is discussed in more detail in,
for example, refs. \cite{AH1} and \cite{Lee}.

\section{Ideal relativistic Bose gas without antiparticles}
To describe an ideal gas of events obeying Bose-Einstein statistics in
the grand canonical ensemble, we use the expression for the number of
events found in \cite{HSP},
\beq
N=V^{(4)}\sum _{k^\mu }n_{k^\mu }=
V^{(4)}\sum _{k^\mu }\frac{1}{e^{(E-\mu -\mu _K\frac{m^2}{2M})/T}-1},
\eeq
where $V^{(4)}$ is the system's four-volume and $m^2 \equiv  -k^2 = - k^\mu
k_\mu ;$
$\mu _K$ is an additional mass potential \cite{HSP}, which arises in the
grand canonical ensemble as the derivative of the free energy with respect to
the value of the dynamical evolution function $K$, interpreted as the invariant
mass of the system.In the kinetic theory \cite{HSP}, $\mu_K$ enters as
a Lagrange multiplier for the equilibrium distribution 
for $K$, as $\mu$ is for $N$, and $1/T$ for $E$.  We shall see, in the
following, how $\mu_K$ plays a fundamental role in determining the
structure of the mass distribution.  In order to simplify subsequent
 considerations, we
shall take it to be a fixed parameter.

 To ensure a
positive-definite value for
$n_{k^\mu },$ the number density of bosons with four-momentum $k^\mu ,$ we
require that
\beq
m-\mu -\mu _K\frac{m^2}{2M}\geq 0.
\eeq
The discriminant for the l.h.s. of the inequality must be nonnegative,
i.e.,
\beq
\mu \leq \frac{M}{2\mu _K}.
\eeq
For such $\mu ,$ (2.2) has the solution
\beq
m_1\equiv \frac{M}{\mu _K}\left( 1-\sqrt{1-\frac{2\mu \mu _K}{M}}\right) \leq
m\leq
\frac{M}{\mu _K}\left( 1+\sqrt{1-\frac{2\mu \mu _K}{M}}\right) \equiv m_2.
\eeq
For small $\mu \mu _K/M,$ the region (2.4) may be approximated by
\beq
\mu \leq m\leq \frac{2M}{\mu _K}.
\eeq
One sees that $\mu _K$ determines an upper
bound of the mass spectrum, in addition to the usual lower bound $m\geq
\mu .$ In fact, small $\mu _K$ admits a very large range of off-shell
mass, and hence can be associated with the presence of strong
interactions \cite{MS}.

Replacing the sum over $k^\mu $ (2.1) by an integral, one obtains for
the density of events per unit space-time volume $n\equiv N/V^{(4)}$
\cite{ind},
\beq
n=\frac{1}{4\pi ^3}\int _{m_1}^{m_2}dm\int _{-\infty }^\infty d\beta
\;\frac{m^3\;\sinh ^2\beta }{
e^{(m\cosh \beta -\mu -\mu _K\frac{m^2}{2M})/T}-1},
\eeq
where $m_1$ and $m_2$ are defined in Eq. (2.4), and we have used the
parametrization \cite{HSS} $$\begin{array}{lcl}
p^0 & = & m\cosh \beta , \\
p^1 & = & m\sinh \beta \sin \theta \cos \phi , \\
p^2 & = & m\sinh \beta \sin \theta \sin \phi , \\
p^3 & = & m\sinh \beta \cos \theta ,
\end{array} $$ $$0\leq \theta <\pi ,\;\;\;0\leq \phi <2\pi ,\;\;\;-\infty
<\beta <\infty .$$

In this paper we shall restrict ourselves to the case of high temperature
alone:
\beq
T>>\frac{M}{\mu _K}.
\eeq
It is then possible to use, for simplicity, the Maxwell-Boltzmann form for the
integrand, and to rewrite (2.6) in the form
\beq
n=\frac{e^{\mu /T}}{4\pi ^3}\int _{m_1}^{m_2}m^3\;dm\int _{-\infty }^
\infty \sinh ^2\beta \;d\beta \;e^{-m\cosh \beta /T}e^{\mu _Km^2/2MT},
\eeq
which reduces, upon integrating out $\beta ,$ to \cite{BH1}
\beq
n=\frac{Te^{\mu /T}}{4\pi ^3}\int _{m_1}^{m_2}dm\;m^2K_1\left( \frac{m}{
T}\right) e^{\mu _Km^2/2MT},
\eeq
where $K_\nu (z)$ is the Bessel function of the third kind (imaginary
argument). Since $\mu \leq m\leq m_2\leq 2M/\mu _K,$
\beq
\frac{\mu _Km^2}{2MT}\leq \frac{\mu _K(2M/\mu _K)^2}{2MT}=\frac{2M}{T
\mu _K}<<1,
\eeq
in view of (2.7), and also
\beq
\frac{\mu }{T}\leq \frac{m}{T}\leq \frac{2M}{T\mu _K}<<1.
\eeq
Therefore, one can neglect the exponentials in Eq. (2.9), and for $K_1(
m/T)$ use the asymptotic formula \cite{AS}
\beq
K_\nu (z)\sim \frac{1}{2}\Gamma (\nu )\left( \frac{z}{2}\right) ^{-\nu },
\;\;\;z<<1.
\eeq
Then, we obtain
\beq
n\cong \frac{T^2}{4\pi ^3}\int _{m_1}^{m_2}dm\;m=\frac{T^2}{2\pi ^3}
\left( \frac{M}{\mu _K}\right) ^2\sqrt{1-\frac{2\mu \mu _K}{M}}.
\eeq
>From this equation, one can identify the high-temperature
mass distribution for the system we are studying, so that now
\beq
\langle m^\ell \rangle =\frac{\int _{m_1}^{m_2}dm\;m^{\ell +1}}
{\int _{m_1}^{m_2}dm\;m}=\frac{2}{\ell +2}\frac{m_2^{\ell +2}-m_1^{\ell
+2}}{m_2^2-m_1^2}.
\eeq
In particular,
\beq
\langle m\rangle =\frac{4}{3}\frac{M}{\mu _K}\left( 1-\frac{\mu \mu _K}{
2M}\right) ,
\eeq
\beq
\langle m^2\rangle =2\left( \frac{M}{\mu _K}\right )^2
\left( 1-\frac{\mu \mu _K}{M}\right) .
\eeq
Extracting the joint distribution for $\beta $ and $m$ from (2.8) in the
same way, we also obtain the average values of the energy and the square of the
energy for high $T.$ The average energy is given by
\beq
\langle E\rangle \equiv \langle m\cosh \beta \rangle \cong \frac{\int
_{m_1}^{m_2}m^4dm\sinh ^2\beta \cosh \beta d\beta e^{-m\cosh \beta /T}}
{\int _{m_1}^{m_2}m^3dm\sinh ^2\beta d\beta e^{-m\cosh \beta /T}}.
\eeq
Integrating out $\beta ,$ one finds
\beq
\langle E\rangle \cong \frac{1}{4T}\frac{\int _{m_1}^{m_2}dm\;m^4[K_3(m/
T)-K_1(m/T)]}{\int _{m_1}^{m_2}dm\;m^2K_1(m/T)}.
\eeq
It is seen, with the help of (2.12), that it is possible to neglect $K_1$
in comparison with $K_3$ in the numerator of (2.18) and obtain, via
(2.12),
\beq
\langle E\rangle \cong \frac{1}{4T}\frac{\int _{m_1}^{m_2}dm\;m^4K_3(m/
T)}{\int _{m_1}^{m_2}dm\;m^2K_1(m/T)}\simeq 2T,
\eeq
in agreement with refs. \cite{HSP,HSS,BH1}. Similarly, one obtains
$$\langle E^2\rangle \equiv \langle m^2\cosh ^2\beta \rangle \cong \frac{
\int _{m_1}^{m_2}m^5dm\sinh ^2\beta \cosh ^2\beta d\beta e^{-m\cosh \beta
/T}}{\int _{m_1}^{m_2}m^3dm\sinh ^2\beta d\beta e^{-m\cosh \beta /T}}$$
\beq
=\frac{\int _{m_1}^{m_2}dm[m^4K_1(m/T)+3Tm^3K_2(m/T)]}{\int _{m_1}^
{m_2}dm\;m^2K_1(m/T)}\cong 3T\frac{\int _{m_1}^{m_2}dm\;m^3K_2(m/T)}{\int
_{m_1}^{m_2}dm\;m^2K_1(m/T)}\simeq 6T^2.
\eeq

Let us assume, as is generally done, that the average $\langle p^\mu p^\nu
\rangle $ has the form
\beq
\langle p^\mu p^\nu \rangle =au^\mu u^\nu +bg^{\mu \nu },
\eeq
where $u^\mu =(1,{\bf 0})$ in the local rest frame. The values of $a$ and
$b$ can then be calculated as follows: for $\mu =\nu =0$ one has
$\langle (p^0)^2\rangle =a-b,$ while contraction of (2.21) with $g^{\mu
\nu }$ gives $-g^{\mu \nu }\langle p_\mu p_\nu \rangle =a-4b.$ The use of
the expressions (2.20) for $\langle (p^0)^2\rangle \equiv \langle E^2
\rangle ,$ and (2.16) for $-g^{\mu \nu }\langle p_\mu p_\nu \rangle
\equiv \langle m^2\rangle $ yields $$\left \{ \begin{array}{rcl}
a-b & = & 6T^2, \\
a-4b & = & 2(\frac{M}{\mu _K})^2\left( 1-\mu \mu _K/M\right) ,
\end{array} \right. $$ so that
\beq
a=8T^2-\frac{2}{3}\left( \frac{M}{\mu _K}\right) ^2
\left( 1-\frac{\mu \mu _K}{M}\right) ,
\eeq
\beq
b=2T^2-\frac{2}{3}\left( \frac{M}{\mu _K}\right) ^2
\left( 1-\frac{\mu \mu _K}{M}\right) .
\eeq
For $T>>M/\mu _K,$ it is possible to take $a\cong 8T^2,$ $b\cong 2T^2,$
and obtain, therefore,
\beq
\langle p^\mu p^\nu \rangle \cong 8T^2u^\mu u^\nu +2T^2g^{\mu \nu }.
\eeq

To find the expressions for
the pressure and energy density in our ensemble, we study the
particle energy-momentum tensor defined by the relation \cite{HSS}
\beq
T^{\mu \nu }(q)=\sum _i\int d\tau \frac{p_i^\mu p_i^\nu }{M/\mu _K}
\delta ^4(q-q_i(\tau )),
\eeq
in which $M/\mu _K$ is the value around which the mass of the bosons
making up the ensemble is distributed, i.e., it corresponds to the
limiting mass-shell value when the inequality (2.3) becomes equality.
Upon integrating over a small space-time volume $\triangle V$ and
taking the ensemble average, (2.25) reduces to \cite{HSS}
\beq
\langle T^{\mu \nu }\rangle =\frac{T_{\triangle V}}{M/\mu _K}n\langle
p^\mu p^\nu \rangle .
\eeq
In this formula $T_{\triangle V}$ is the average passage interval in
$\tau $ for the events which pass through the small (typical) four-volume
$\triangle V$ in the neighborhood of the $R^4$-point. The four-volume
$\triangle V$ is the smallest that can be considered a macrovolume in
representing the ensemble. Using the standard expression
\beq
\langle T^{\mu \nu }\rangle =(p+\rho )u^\mu u^\nu +pg^{\mu \nu },
\eeq
where $p$ and $\rho $ are the particle pressure and energy density,
respectively, we obtain
\beq
p\equiv p(\mu )=\frac{T_{\triangle V}}{\pi ^3}\frac{M}{\mu _K}\sqrt{1-\frac{2\mu
\mu _K}{M}}T^4,\;\;\;\rho =3p.
\eeq
To interpret these results we calculate the particle number density per
unit three-volume. The particle four-current is given by the
formula \cite{HSS}
\beq
J^\mu (q)=\sum _i\int d\tau \frac{p^\mu _i}{M/\mu _K}\delta ^4(q-q_i(
\tau )),
\eeq
which upon integrating over a small space-time volume and taking the
average reduces to
\beq
\langle J^\mu \rangle =\frac{T_{\triangle V}}{M/\mu _K}n\langle p^\mu
\rangle ;
\eeq
then
\beq
N_0\equiv \langle J^0\rangle =\frac{T_{\triangle V}}{M/\mu _K}n
\langle E\rangle ,
\eeq
so that
\beq
N_0\equiv N_0(\mu )=\frac{T_{\triangle V}}{\pi ^3}\frac{M}{\mu
_K}\sqrt{1-\frac{2\mu
\mu _K}{M}}T^3,
\eeq
and we recover the ideal gas law
\beq
p=N_0T.
\eeq

Since, in view of (2.4), $$\frac{2M}{\mu _K}\sqrt{1-\frac{2\mu \mu _K}{
M}}=\triangle m$$ is a width of the mass distribution around the
value $M/\mu _K,$ Eqs. (2.28),(2.32) can be rewritten as
\bqry
p & = & \frac{T_{\triangle V}\triangle m}{2\pi ^3}T^4,\;\;\;
\rho \;=\;3p, \NL
N_0 & = & \frac{T_{\triangle V}\triangle m}{2\pi ^3}T^3.
\eqry
In ref. \cite{glim} we obtained the formulas for thermodynamic variables, under
the assumption of narrow  mass width, which depend on $T_{\triangle V}
\triangle m$ as well; the requirement that these results coincide with those of
the usual on-shell theories implies the relation\footnote{In c.g.s. units, this
relation has a factor $\hbar /c^2$ on the right hand side.}
\beq
T_{\triangle V}\triangle m=2\pi .
\eeq
One can understand this relation, up to a numerical factor, in terms of
the uncertainty principle (rigorous in the $L^2(R^4)$ quantum theory)
$\triangle E\cdot \triangle t\stackrel{>}{\sim }1/2.$ Since the time
interval for the particle to pass the volume $\triangle V$ (this smallest
macroscopic volume is bounded from below by the size of the wave packets)
$\triangle t\cong E/M \;\triangle \tau ,$ and the dispersion of $E$ due
to the mass distribution is $\triangle E\sim m\triangle m/E,$ one obtains
a lower bound for $T_{\triangle V}\triangle m$ of order unity.

Thus, with (2.35) holding, the formulas (2.34)  reduce to
\bqry
p & = & \frac{T^4}{\pi ^2},\;\;\;\rho \;=\;3p, \\
N_0 & = & \frac{T^3}{\pi ^2},
\eqry
which are the standard expressions for high temperature \cite{Kolb}.
The formulas for characteristic thermodynamic quantities and the equation
of state for a relativistic gas of off-shell events have the same form as those
of the relativistic gas of on-shell particles. They coincide with them (under
the condition (2.35)) in the narrow mass shell limit, except for the expression
for the average energy which takes the value
$2T$ in the relativistic gas of events, in contrast to $3T,$ as for the
high-temperature limit of the usual theory \cite{Pauli}. Experimental
measurement of average energy at high temperature can, therefore, affirm (or
negate) the validity of the off-shell theory. There seems to be no empirical
evidence which distinguishes between these results at the present time. The
quantity
$\sigma =M_0c^2/k_BT,$ a parameter which distinguishes the relativistic
from the nonrelativistic regime (see, e.g., \cite{HSP}) is very large for
$M_0$ of the order of the pion mass, at ordinary temperatures; the
ultrarelativistic limit corresponding to $\sigma $ small becomes a
reasonable approximation for $T\stackrel{>}{\sim }10^{12}$ K.

\section{Antiparticles and condensation}
The introduction of antiparticles into the theory as the CPT conjugate of
negative energy events leads, by application
of the arguments of Haber and Weldon \cite{HW1}, or Actor \cite{Act}, to
a change in sign of $\mu $ in the distribution function for
antiparticles. We therefore write down the following relation which
represents the analog of the formula (1.2): \footnote{As for the
nonrelativistic theory, the ``free'' distribution functions describe
quasiparticles in a form which takes interactions into account entering
through the chemical potential. By definition, good
quasiparticles are not frequently emitted or absorbed; we therefore
consider the (quasi-) particles and antiparticles as two species. Since the
particle number is determined by the derivative of the free energy with
respect to the chemical potential, $\mu $ must change sign for the
antiparticles \cite{HW1}. Similarly, the average mass (squared) is
obtained by the derivative with respect to $\mu _K$ \cite{HSP}; since the mass
(squared) of the antiparticle is positive, $\mu _K$ does not change sign.}
\beq
N=V^{(4)}\sum _{k^\mu }\left[ \frac{1}{e^{(E-\mu -\mu _K\frac{m^2}{2M})
/T}-1}-\frac{1}{e^{(E+\mu -\mu _K\frac{m^2}{2M})/T}-1}\right] .
\eeq
With respect to the determination of the sign of the second term, let us
 consider a space-time picture in which we have many world lines, generated by
events moving monotonically in the positive $t$ direction.  The addition of a
particle-antiparticle pair which annihilates corresponds to the addition of a
world line which is generated by an event initially moving in the positive
direction of time to some upper bound, $t_0$, where annihilation takes place,
and returning in the negative direction of time.  At times later than $t_0$ the
total particle number is unaffected.  At times earlier than $t_0$, a particle
and antiparticle are added to the total paticle number.  Since, as also assumed
by Haber and Weldon \cite{HW1}, the total particle number is a conserved
quantity, the antiparticle trajectory must be counted with a sign oppostie to
that of the particle trajectory.  The second term in (3.1), counting
antiparticles, must therefore carry a negative sign.  We require that both
$n_{k^\mu }$ terms in Eq. (3.1) be positive definite. In this way we obtain the
two quadratic inequalities,
\bqry
m-\mu -\mu _K\frac{m^2}{2M} & \geq  & 0, \NL
m+\mu -\mu _K\frac{m^2}{2M} & \geq  & 0,
\eqry
which give the following relation representing the nonnegativeness of
the corresponding discriminants:
\beq
-\frac{M}{2\mu _K}\leq \mu \leq \frac{M}{2\mu _K}.
\eeq
It then follows that we must consider the intersection of
the ranges of validity of the two inequalities (3.2). Indeed, if each
inequality is treated separately, there would be some values of $m$ for
which one and not another would be physically acceptable. One finds the
bounds of this intersection region by solving these inequalities, and
obtains\footnote{This is actually the solution of one of the inequalities (3.2)
(the most restrictive), depending on the sign of $\mu .$}
\beq
\frac{M}{\mu _K}\left( 1-\sqrt{1-\frac{2|\mu |\mu _K}{M}}\right)\leq m
\leq \frac{M}{\mu _K}\left(1+\sqrt{1-\frac{2|\mu |\mu _K}{M}}\right) ,
\eeq
which for small $|\mu |\mu _K/M$ reduces, as
in the no-antiparticle case (2.5), to
\beq
|\mu |\leq m\leq \frac{2M}{\mu _K}.
\eeq

Replacing the summation in (3.1) by integration, we obtain a
formula for the number density:
\beq
n=\frac{1}{4\pi ^3}\int _{m_1}^{m_2}\!\!m^3dm\int _{-\infty }^\infty
\!\!\sinh ^2\beta d\beta \left[
\frac{1}{e^{(m\cosh \beta -\mu -\mu _K\frac{m^2}{2M})/T}-1}-
\frac{1}{e^{(m\cosh \beta +\mu -\mu _K\frac{m^2}{2M})/T}-1}\right] ,
\eeq
where $m_1$ and $m_2$ are defined in Eq. (3.4), which for large $T$
reduces, as above, to $$n=\frac{e^{\mu /T}-e^{-\mu /T}}{4\pi ^3}T\int
_{m_1}^{m_2}dm\;m^2K_1\left( \frac{m}{T}\right) e^{\mu _Km^2/2MT}.$$
Now, using the estimates (2.10),(2.11), and $\sinh (\mu /T)\cong \mu /T$
for $\mu /T<<1,$ we obtain (in place of (2.13)) the net event charge
\beq
n=\frac{1}{\pi ^3}\left( \frac{M}{\mu _K}\right) ^2
\sqrt{1-\frac{2|\mu |\mu _K}{M}}\mu T.
\eeq
The pressure and energy density are obtained by the sum particle and
antiparticle contributions (proportional to $\exp (\pm \mu /T)$, with the
number density (3.7).  To second order in $(\mu / T)^2 $, one finds
\bqryn
p & = & 2p(|\mu | ), \\
\rho  & = & 2\rho (|\mu |),
\eqryn
where $p(\mu )$ and $\rho (\mu )$ are given by (2.28) with $\mu $
replaced by $|\mu |.$ On the other hand, from (2.31) and (3.7), one finds
\beq
N_0=2\frac{T_{\triangle V}}{\pi ^3}\frac{M}{\mu _K}\sqrt{
1-\frac{2|\mu |\mu _K}{M}}\mu T^2,
\eeq
where the factor of $2\mu /T,$ as compared to (2.32), arises from the
{\it difference} between the factors $\exp (\pm \mu /T)$ (the sign of $\mu $
indicates whether particles or antiparticles predominate). One then obtains the
following expressions for the Bose gas including both particles and
antiparticles\footnote{If we did not neglect indistinguishability of bosons at
high temperature, we would obtain, instead of (2.37)
\cite{glim}, $N_0=\frac{T^3}{\pi ^2}Li_3(e^{\mu /T}),$ where $Li_\nu (z)
\equiv \sum _{s=1}^\infty z^s/s^\nu $ is the polylogarithm \cite{Prud},
so that, for the system including both particles and antiparticles,
$N_0=\frac{T^3}{\pi ^2}[Li_3(e^{\mu /T})-Li_3(e^{-\mu /T})].$ It then
follows from the properties of the polylogarithms \cite{Prud} that, for
$x\equiv |\mu |/T<<1,$ $Li_3(e^x)-Li_3(e^{-x})\cong \frac{\pi ^2}{3}x,$
so that, we would obtain, instead of (3.10), $N_0=\mu T^2/3,$ which
coincides with Haber and Weldon's Eq. (1.5).} (here, $\triangle m$ is not
necessarily small):
\bqry
p & = & \frac{T_{\triangle V}\triangle m}{2\pi }\frac{2T^4}{\pi ^2},\;\;\;\rho
\;=\;3p,
\\ N_0 & = & \frac{T_{\triangle V}\triangle m}{2\pi }\frac{2T^2}{\pi ^2}\mu .
\eqry

We now wish to show that the dynamical properties of the current, which
follow from the relativistic canonical equations of motion, are consistent with
the thermodynamic relation
\beq
N_0=\frac{N}{V},
\eeq
where $N$ is the number of bosons in a three-dimensional box of volume
$V.$ Since the event number density $n$ is, by definition,
$$n=\frac{N}{V^{(4)}}=\frac{N}{V\triangle t},$$ where $\triangle t$ is
the (average) extent of the ensemble along the $q^0$-axis (as in our
discussion after (2.35)), one has
\beq
N_0=n\triangle t.
\eeq
The equation of motion (1.9) for $q^0$ (with $M/\mu _K,$ the central
value of the mass distribution, instead of $M,$
which corresponds to a change of scale parameter in the expression (1.8)
for the generalized Hamiltonian $K),$ $$\frac{dq_i^0}{d\tau }=\frac{p_i^0
}{M/\mu _K},$$ upon averaging over the whole ensemble, reduces to
\beq
\frac{\triangle t}{T_{\triangle V}}=\frac{\langle E\rangle }{M/\mu _K},
\eeq
where $T_{\triangle V}$ is the average passage interval in $\tau $ used
in previous consideration. Then, in view of (3.12),(3.13), one obtains
the equation (2.31).

\subsection{Relativistic Bose-Einstein condensation}
Since in the particle-antiparticle case, $N_{{\rm rel}}\equiv N-\bar{N},$
where $N$ and $\bar{N}$ are the numbers of particles and antiparticles,
respectively, is a conserved quantity, according to the arguments of
Haber and Weldon \cite{HW1} pointed out in
Section 1, and our discussion above, $N_0=N_{{\rm rel}}/V$ is also a conserved
 quantity, so that
it makes sense to talk of $|N_{{\rm rel}}|$ bosons in a spatial box of
the volume $V.$ Therefore, in Eq. (3.10) $N_0$ is a conserved quantity,
so that, the dependence of $\mu $ on temperature is defined by (we
assume that $N_0$ is continuous at the phase transition)
\beq
\mu =\frac{2\pi }{T_{\triangle V}\triangle m}\frac{\pi ^2N_0}{2T^2}.
\eeq

For $T$ above some critical temperature, one can always find $\mu $
satisfying (3.3) such that the relation (3.14) holds; no such $\mu $
can be found for $T$ below the critical temperature. The value of the
critical temperature is defined by putting $|\mu |=M/2\mu _K$ in (3.14). In the
narrow mass-shell limit, inserting (2.35), one obtains
\beq
T_c=\pi \sqrt{\frac{|N_0|}{M/\mu _K}}.
\eeq
For $|\mu |=M/2\mu _K,$ the width of the mass distribution is zero, in
view of (3.4), and hence the ensemble approaches a distribution sharply
peaked at the mass-shell value $M/\mu _K.$ The fluctuations $\delta m=
\sqrt{\langle m^2\rangle -\langle m\rangle ^2}$ also vanish. Indeed, as
follows from (2.15),(2.16) with $\mu $ replaced by $|\mu |,$ and
(3.14),(3.15),
\beq
\delta m=\frac{M}{3\mu _K}\sqrt{2-\left( \frac{T_c}{T}\right) ^2-
\left( \frac{T_c}{T}\right) ^4},
\eeq
so that, at $T=T_c,$ $\delta m=0.$ It follows from (3.16) that for $T$
in the vicinity of $T_c$ $(T\geq T_c),$
\beq
\delta m\simeq \frac{M}{3\mu _K}\sqrt{\frac{6}{T_c}}\sqrt{T-T_c},
\eeq
as for a second order phase transition, for which fluctuations go to zero
smoothly.

We note that Eqs. (2.36),(2.37) do not contain explicit dependence on the
chemical potential, and hence no phase transition is induced. In fact, at
lower temperature (or small $\mu _K)$ one or the other of the particle or
antiparticle distribution dominates, and one returns to the case of the
high-temperature strongly interacting gas \cite{BHS1}. The remaining
phase transition is the usual low-temperature Bose-Einstein condensation
discussed in the textbooks.

One sees, with the help of (3.4), that the expression for $n$ (3.7) can
be rewritten as
\beq
n=\frac{1}{2\pi ^3}\frac{M}{\mu _K}\triangle m\;\mu T;
\eeq
since at $T=T_c,$ $\triangle m=0,$ it follows that $n=0$ at all
temperatures below $T_c.$ Therefore, the behaviour of an
ultrarelativistic Bose gas including both particles and antiparticles,
which is governed by the relation (3.14), can be thought of as a special
type of Bose-Einstein condensation to a ground state with $p^\mu p_\mu
=-(M/\mu _K)^2$ (this ground state occurs with zero weight in the
integral (3.6)). In such a formulation, every state with temperature
$T>T_c,$ given by Eq. (3.6), should be considered as an {\it off-shell}
excitation of the on-shell ground state. At $T=T_c,$ all such excitations
freeze out and the distribution becomes strongly peaked at a definite
mass, i.e., the system undergoes a phase transition to the on-shell
sector. Note that, for $n=0,$ Eq. (3.12) gives $\triangle t=\infty.$ Then,
since $\langle E\rangle \sim T,$ one obtains from (3.13) that $T_{
\triangle V}=\infty$ (this relation can be also obtained from (2.35) for
$\triangle m=0),$ which means that in the mean, all the events become particles.

 As the distribution function enters
the on-shell phase at $T=T_c,$ the underlying off-shell theory describes
fluctuations around the sharp mean mass. This phenomenon provides a
mechanism, based on equilibrium statistical mechanics, for understanding
how the general off-shell theory is constrained to the neighborhood of a
sharp universal mass shell for each particle type. At temperatures below
$T_c,$ the results of the theory for the main thermodynamic quantities
coincide with those of the usual on-shell theories.

In order that our considerations be valid, the relation
$T_c>>M/\mu _K$ must hold; this relation reduces, with (3.15), to
\beq
\Big| N_0\Big| >>\frac{1}{\pi ^2}\left( \frac{M}{\mu _K}\right) ^3.
\eeq
For $M/\mu _K\sim m_\pi \simeq 140$ MeV, this inequality yields $N_0>>
3\cdot 10^5\;{\rm MeV}^3.$ Taking $N_0\sim 5\cdot 10^6\;{\rm MeV}^3,$
which corresponds to temperature $\sim 350$ MeV, in view of (2.37),
one gets $T_c\sim 550\;{\rm MeV}\simeq 4m_\pi .$

If $\mu _K$ is very small, it is difficult to satisfy (3.19) and the
possibility of such a phase transition may disappear. This case
corresponds, as noted above, to that of strong interactions and is
discussed in succeeding paper \cite{BHS1}.

\section{Concluding remarks}
We have considered the ideal relativistic Bose gas within the framework
of a manifestly covariant relativistic statistical mechanics, taking
account of antiparticles. We have shown that in such a
particle-antiparticle system at some critical temperature $T_c$ a
special type of relativistic Bose-Einstein condensation sets in, which
corresponds to a phase transition from the sector of relativistic mass
distributions to a sector in which the boson mass distribution peaks at a
definite mass. The results which can be computed from the latter
coincide with those obtained in a high-temperature limit of the usual
on-shell relativistic theory.

The relativistic Bose-Einstein condensation in particle-antiparticle
system considered in the present paper can represent (as for the
Galilean limit $c\rightarrow \infty $ \cite{glim}) a possible mechanism
of acquiring a given sharp mass distribution by the particles of the system,
as a phase transition between the corresponding sectors of the theory. Since
this phase transition can occur at an ultrarelativistic temperature, it
might be relevant to cosmological models. The relativistic Bose-Einstein
condensation considered in the present paper may also have properties
which could be useful in the study of relativistic boson stars
\cite{Jet}. These and the other aspects of the theory are now under
further investigation.

The extension and generalization of Bose-Einstein condensation to curved
space-times and space-times with boundaries, for which the work
reported here may have constructive application,  has also been the subject of
much study. The non-relativistic Bose gas in the Einstein static universe
was treated in ref. \cite{Al}. The generalization to relativistic scalar
fields was given in refs. \cite{SP,PZ}. The extension to higher
dimensional spheres was given in ref. \cite{Shi}. Bose-Einstein
condensation on hyperbolic manifolds \cite{CV}, and in the Taub universe
\cite{Hua1} has also been considered. More recently, by calculating the
high-temperature expansion of the thermodynamic potential when the
boundaries are present, Kirsten \cite{Kir} examined Bose-Einstein
condensation in certain cases. Later work of Toms \cite{Toms} showed how
to interpret Bose-Einstein condensation in terms of symmetry breaking, in
the manner of flat space-time calculations \cite{HW2,Kap}. The most
recent study by Lee {\it et al.} \cite{LKK} showed how interacting scalar
fields can be treated. Bose-Einstein condensation for self-interacting
complex scalar fields was considered in ref. \cite{KT}. It is to be hoped
that the techniques developed here can contribute to the development of this
subject as well.

\section*{Acknowledgements}
  We wish to thank the Los Alamos National Laboratory
and the Ilya Prigogine Center for Studies in Statistical Mechanics and Complex
Systems at Austin, Texas for their hospitality in enabling us to collaborate
closely in the final stages of this work.

\end{document}